\documentclass[fleqn,twoside]{article}
\usepackage{espcrc2}
\usepackage{epsfig}

\usepackage{graphicx}
\usepackage[figuresright]{rotating}

\newcommand{\AmS}{{\protect\the\textfont2
  A\kern-.1667em\lower.5ex\hbox{M}\kern-.125emS}}

\title{The joint extraction of $m_s$ and $V_{us}$ from hadronic $\tau$
decay data}

\author{K. Maltman\address{Department of Math and Stats, York Univ., 
Toronto, ON M3J 1P3, Canada and
CSSM, Univ. of Adelaide, Adelaide, SA, 5005, Australia}
\thanks{e-mail: kmaltman@yorku.ca. Work supported by a grant from
the Natural Sciences and Engineering Research Council of Canada},
C.E. Wolfe\address{Department of Physics and Astronomy, York Univ., 
Toronto, ON M3J 1P3, Canada}}
       
\begin{document}

\begin{abstract}
We discuss the simultaneous determination of $m_s$ and $V_{us}$ 
from flavor-breaking hadronic $\tau$ decay sum rules, focussing
on weight choices designed to better control problems associated with the 
slow convergence of the relevant integrated $D=2$ OPE series.
The results are found to display
improved stability and consistency relative to those of
conventional analyses based on the ``$(k,0)$ spectral weights''. 
The results for $m_s$ agree well with those of recent strange scalar 
sum rule and strange pseudoscalar sum rule and lattice analyses. Results
for $V_{us}$ agree within errors with those of lattice-input-based
$\Gamma [K_{\mu 2}] /\Gamma [\pi_{\mu 2}]$ and $K_{\ell 3}$
analyses. Very significant error reductions are shown
to be expected, especially for $V_{us}$, once improved strange
spectral data from the B factory experiments becomes available.
\vspace{1pc}
\end{abstract}

\maketitle

\section{BACKGROUND}
Measurements of the inclusive hadronic $\tau$ decay distributions for processes
mediated by the flavor $ij=ud,us$ vector (V) or axial vector (A) currents 
yield kinematically weighted linear combinations 
of the spectral functions, $\rho^{(J)}_{V/A;ij}$, of the spin $J=0$ and $1$ 
parts of the relevant current-current correlators, $\Pi^{(J)}_{V/A;ij}$.
With $R_{V/A;ij}\, \equiv\, \Gamma [\tau^- \rightarrow \nu_\tau
\, {\rm hadrons}_{V/A;ij}\, (\gamma )]/
\Gamma [\tau^- \rightarrow \nu_\tau e^- {\bar \nu}_e (\gamma)]$,
one has, explicitly~\cite{bnpetc},
\begin{eqnarray}
&&{\frac{R_{V/A;ij}}{\left[ 12\pi^2\vert V_{ij}\vert^2 S_{EW}\right]}}
=\int^{1}_0\, dy_\tau \,
\left( 1-y_\tau\right)^2 \nonumber\\
&&\quad\left[ \left( 1 + 2y_\tau\right)
\rho_{V/A;ij}^{(0+1)}(s) - 2y_\tau \rho_{V/A;ij}^{(0)}(s) \right]
\label{basictaudecay}
\end{eqnarray}
where $y_\tau =s/m_\tau^2$, $V_{ij}$ is the
flavor $ij$ CKM matrix element, $S_{EW}$ is 
a short-distance electroweak correction, and the superscript
$(0+1)$ denotes the sum of $J=0$ and $J=1$ contributions.
The absence of kinematic singularities in the correlators
corresponding to the spectral function combinations 
in Eq.~(\ref{basictaudecay}) allows the spectral integrals
to be re-written using the basic finite energy sum rule
(FESR) relation. For such correlators $\Pi$, with associated
spectral functions $\rho$, and $w(s)$ analytic, this relation has the form
\begin{equation}
\int_0^{s_0}ds w(s) \rho (s) =\, {\frac{-1}{2\pi i}}\,
\oint_{\vert s\vert =s_0}ds w(s) \Pi (s)\ .
\label{fesrbasic}\end{equation}
Analogous FESR's, with spectral integral sides denoted $R_{V/A;ij}^{(k,m)}$,
obtained from $R_{V/A;ij}$ by rescaling the integrand with 
$(1-y_\tau )^ky_\tau^m$ before integration, are called the 
``$(k,m)$ spectral weight sum rules''. Similar spectral integrals and 
sum rules can be constructed for general non-spectral weights $w(s)$,
for either $\Pi^{(0+1)}_{V/A;ij}(s)$ or $s\Pi^{(0)}_{V/A;ij}(s)$, and
for any $s_0<m_\tau^2$. The corresponding spectral integrals
are denoted generically by $R^w_{ij}(s_0)$ in what follows.
For ``inclusive'' sum rules (those with both $J=0$ and
$J=0+1$ spectral contributions) the purely $J=0$
contribution will be referred to as ``longitudinal''.

$V_{us}$ and/or $m_s$ are then to be extracted using flavor-breaking 
differences, $\delta R^w(s_0)$, defined by
\begin{equation}
\delta R^w(s_0)\, =\, {\frac{R^w_{ud}(s_0)}{\vert V_{ud}\vert^2}}
\, -\, {\frac{R^w_{us}(s_0)}{\vert V_{us}\vert^2}}\ .
\label{tauvusbasicidea}\end{equation}
Since $\delta R^w(s_0)$ vanishes in the $SU(3)_F$ limit, its OPE 
representation, $\delta R^w_{OPE}(s_0)$, begins at dimension $D=2$. 
The $D=2$ term is proportional to $m_s^2$. Experimental determinations of
$\delta R^w(s_0)$ over a range of $s_0$ and $w(s)$
allow a fit for $m_s$ and/or $V_{us}$ to be performed, so long as
$s_0$ is large enough that the OPE representation can be reliably
employed~\cite{taumsrefs,km00,longposviol,pichetalvus,kmtau0204,kmcwvus06}.
The approach is especially well-suited to the determination of
$V_{us}$~\cite{pichetalvus,kmcwvus06} since the smallness of $m_s$
makes the integrated $D=2$ OPE contributions at scales 
$\sim 2-3\ {\rm GeV}^2$, and hence also the corresponding flavor-breaking 
spectral integral differences, much smaller than the individual flavor 
$ud$ and $us$ spectral integral terms (typically at the few to several 
percent level). Explicitly, since Eq.~(\ref{tauvusbasicidea}) implies
\begin{equation}
\vert V_{us}\vert \, =\, \sqrt{ {\frac{R^w_{us}(s_0)}{ 
\left[ {\frac{R^w_{ud}(s_0)}{\vert V_{ud}\vert^2}}
\, -\, \delta R^w_{OPE}(s_0)\right]}}}\ ,
\label{tauvussolution}\end{equation}
it follows that an uncertainty $\Delta \left(\delta R^w_{OPE}(s_0)\right)$
in $\delta R^w_{OPE}(s_0)$ produces a fractional uncertainty
$\simeq \Delta \left(\delta R^w_{OPE}(s_0)\right)/2\, R^w_{ud}(s_0)$
in $\vert V_{us}\vert$ which is {\it much} smaller than that on 
$\delta R^w_{OPE}(s_0)$ itself~\cite{pichetalvus}. Moderate precision for 
$\delta R^w_{OPE}(s_0)$ thus suffices for high precision on 
$\vert V_{us}\vert$, provided experimental errors can be brought under control.

We report here on a combined extraction of $m_s$ and $\vert V_{us}\vert$ 
employing existing spectral data. The $ud$ data, especially the V+A sum, 
are already quite precise~\cite{cleoud,alephud,alephfinalspec,opalud}.
The $us$ data~\cite{alephfinalspec,alephus99,cleous0305,opalus04}, however,
suffer from low statistics and have very sizeable errors above the $K^*$. 
Ongoing analyses at BABAR and BELLE will very significantly reduce
the $us$ errors in the near future. The focus here will be on bringing
uncertainties on the theoretical (OPE) side of the analysis under
better control, in particular those associated with the 
slower-than-previously-anticipated convergence of the relevant $D=2$ OPE 
series~\cite{bck05}.

\section{OPE COMPLICATIONS}
The first major problem on the OPE side is the very bad behavior of the 
integrated longtitudinal $D=2$ OPE series which shows no sign of 
converging at any kinematically accessible scale~\cite{bck05,longconv}, and,
even worse, for {\it all} truncation schemes used
in the literature, badly violates spectral-positivity-based
constraints~\cite{longposviol,longposviolfootnote}.
Inclusive analyses employing the longitudinal OPE representation 
are thus untenable.

Fortunately, for a combination
of chiral and kinematic reasons, the longitudinal $D=2$ OPE problem is
easily handled phenomenologically. Apart from the $\pi$ and
$K$ pole terms, longitudinal spectral contributions
vanish in the $SU(3)_F$ limit and are doubly-chirally suppressed
away from it. This double chiral suppression is preserved in the
ratio of integrated non-pole to pole contributions 
because the longitudinal kinematic weight in Eq.~(\ref{basictaudecay}) 
has essentially the same value at the $K$ pole as in
the region of excited strange scalar and pseudoscalar (PS) resonances.
The small residual non-pole $us$ PS and scalar contributions
can, moreover, be well-constrained phenomenologically~\cite{km00}, the 
former via a sum rule analysis of the $us$ PS channel~\cite{mkps}, 
the latter by dispersive single- or coupled-channel
$K\pi$-scattering-data-based dispersive analyses~\cite{jopss,msssnew}. 
With the dominant $\pi$ and $K$ pole contributions already accurately
known, a bin-by-bin subtraction of the longitudinal 
contributions to the experimental distribution, and hence
a determination of the $(0+1)$ spectral function, can be performed,
allowing FESR's not afflicted by the longitudinal $D=2$ OPE problem to 
be constructed. 

Since the $us$ scalar and PS spectral ``models'' 
used for the longitudinal subtraction correspond, 
via scalar and PS sum rules~\cite{mkps,msssnew,sumrulems}, to 
values of $m_s$ in excellent agreement with recent $N_f=2+1$ lattice 
results~\cite{mslattice}, significantly larger non-pole contributions
are ruled out. The residual non-pole longitudinal subtractions 
are thus certainly small, and very well under control
at the level required for $(0+1)$ FESR analyses. We thus
focus, in what follows, on sum rules involving 
the flavor-breaking combination
\begin{equation}
\Delta\Pi (s)\, \equiv\, \Pi_{V+A;ud}^{(0+1)}(s)\, -\, 
\Pi_{V+A;us}^{(0+1)}(s)\ .
\label{correlatorchoice}\end{equation}

The $D=2$ OPE contribution to $\Delta\Pi$ is known up to
$O(\alpha_s^3)$ and given by~\cite{bck05}
\begin{eqnarray}
&&\left[\Delta\Pi (Q^2)\right]^{OPE}_{D=2}\, =\, {\frac{3}{2\pi^2}}\,
{\frac{\bar{m}_s}{Q^2}} \left[ 1+2.333 \bar{a}+\right. \nonumber\\
&&\left. 19.933 \bar{a}^2 +208.746 \bar{a}^3+(2378\pm 200)\bar{a}^4
\right]
\label{d2form}\end{eqnarray}
where $\bar{a}=\alpha_s(Q^2)/\pi$ and $\bar{m}_s=m_s(Q^2)$, 
with $\alpha_s(Q^2)$ and $m_s(Q^2)$ the running coupling and strange quark 
mass in the $\overline{MS}$ scheme. The $O(\bar{a}^4)$ coefficient
is an estimate obtained using approaches which accurately predicted 
the $O(\bar{a}^3)$ coefficient in Eq.~(\ref{d2form}) 
and $n_f$-dependent $O(\bar{a}^3m_q^2)$ coefficients of the electromagnetic 
current correlator in advance of their explicit calculation~\cite{bckzin04}.
Since independent high-scale determinations of $\alpha_s(M_Z)$~\cite{pdg06}
correspond, after 4-loop running and matching~\cite{cks97,betagamma4},
to $\bar{a}(m_\tau^2)\simeq 0.10-0.11$, the convergence of this
series, at the spacelike point on the contour $\vert s\vert =s_0$,
is marginal at best, even at the highest scales accessible in hadronic 
$\tau$ decay. Although $\vert \alpha_s(Q^2)\vert$ {\it does} decrease as one 
moves around the contour away from the spacelike point, allowing 
improvement in the convergence through judicious choices of weight, $w(s)$, 
one must expect to find very slow convergence 
of the integrated $D=2$ series for those $w(s)$ not chosen specifically with 
this improvement criterion in mind. The $(k,0)$ spectral weights,
$w^{(k,0)}(y)=(1+2y)(1-y)^{k+2}$, with $y=s/s_0$, are highly non-optimal in 
this regard, since $\vert 1-y\vert = 2\vert sin(\phi /2)\vert$ (with $\phi$ 
the angular position measured counterclockwise from the timelike point) 
is peaked precisely in the spacelike direction. Slow convergence, 
deteriorating with increasing $k$, is thus expected for the integrated 
$D=2$ series of the $(k,0)$ spectral weights. The results of Table I of 
Ref.~\cite{bck05} bear out this expectation~\cite{footnote1}.

Several estimates of the integrated $(0+1)$ $D=2$ OPE
truncation uncertainty have been considered in the literature: 
the size of the last term kept, the level of residual scale 
dependence, the difference between the truncated correlator and Adler 
function evaluations, and various combinations thereof.
The slow convergence of the integrated series can make it difficult
to be sufficiently conservative. E.g., the quadrature sum
of the last term size plus residual scale dependence version of
the $O(\bar{a}^3)$ Adler function truncation 
uncertainty~\cite{pichetalvus}, yields a result $\sim 2.5$ times smaller than 
the actual difference between the $O(\bar{a}^3)$-truncated Adler function and 
$O(\bar{a}^4)$-truncated correlator results~\cite{kmcwvus06}.

Since, due to the growth of $\alpha_s$ with decreasing scale,
higher order terms are relatively more important at lower scales,
premature truncation of a slowly converging series will induce
an unphysical $s_0$-dependence in extracted, nominally $s_0$-independent 
quantities. With polynomial weights, $w(y)=\sum_m c_my^m$
(for which integrated $D=2N+2$ OPE contributions not suppressed by 
additional powers of $\alpha_s$ scale as $c_N/s_0^N$)
such unphysical $s_0$-dependence can also result if
higher $D$ contributions which might in principle be present 
are incorrectly assumed negligible and omitted from the analysis. 
The absence of phenomenological input for the values of the relevant
$D>6$ condensates makes such omission most dangerous for those
$w(y)$ having large coefficients $c_m$, with $m>2$,
where such unknown $D>6$ contributions are potentially enhanced.
The $(2,0)$, $(3,0)$ and $(4,0)$ 
spectral weights, $w^{(2,0)}(y)=1-2y-2y^2+8y^3-7y^4+2y^5$,
$w^{(3,0)}(y)=1-3y+10y^3-15y^4+9y^5-2y^6$,
and $w^{(4,0)}(y)=1-4y+3y^2+10y^3-25y^4+24y^5-11y^6+2y^7$
provide examples of weights having such large higher order coefficients.

In view of the above discussion, 
$s_0$-stability tests are crucial to establishing the reliability of 
any theoretical error estimate. 
Failure to find a stability window in $s_0$ or, if not a stability window,
then at least a window within which the observed instability
is safely smaller than the estimated theoretical uncertainty,
is a clear sign of an insufficiently conservative error.

\section{RESULTS AND DISCUSSION}
We restrict our attention, in what follows,
to the V+A spectral combination, to weights satisfying $w(s=s_0)=0$, and 
to scales $s_0>2\ {\rm GeV}^2$, 
all of which serve to strongly suppress possible residual OPE breakdown 
effects~\cite{kmtau0204,kmfesr,cdgmudvma}.

The form of the known $D=4,6$ contributions to
$\left[\Delta\Pi (Q^2)\right]_{OPE}$ may be found in 
Ref.~\cite{bnpetc}. Standard values for the required
OPE input parameters are employed (for details,
see Ref.~\cite{kmcwinprep}). Our central $D=2$ determinations
employ the contour-improved (CIPT) prescription~\cite{ciptbasic} for the
RG-improved correlator difference $\left[\Delta\Pi\right]^{OPE}_{D=2}$.
An alternate CIPT evaluation, employing the truncated, RG-improved 
Adler function, provides one measure of the truncation
uncertainty. (The two versions are equal to all orders but
differ at $O(\bar{a}^{N+1})$ and higher
when both are truncated at $O(\bar{a}^N)$.)
Exact solutions corresponding to the 4-loop-truncated $\beta$ and 
$\gamma$ functions~\cite{betagamma4} are used 
for the running of $\bar{a}$ and $\bar{m}_s$.

For the spectral integrals we employ the ALEPH $ud$~\cite{alephud} 
and $us$~\cite{alephus99} data, for which both data and covariance matrices are
publicly available. A small global renormalization of the $ud$
data is required to reflect minor changes in the $e$, $\mu$ and total
strange branching fractions since the original ALEPH publication. 
Global mode-by-mode rescalings of the strange exclusive distributions
have also been performed to bring the original ALEPH branching fractions
into agreement with current world average (PDG06) values~\cite{chen01}. 
Errors on the $K$ and $\pi$ 
pole contributions have also been reduced by using the more precise 
values implied by $\Gamma [\pi_{\mu 2}]$ and $\Gamma [K_{\mu 2}]$. 
The resulting errors on the $ud$ and $us$ spectral integrals
are at the $\sim 0.5\%$ and $\sim 3-4\%$ levels, respectively.
The $us$ errors will be drastically reduced by results from BABAR and BELLE.

\subsection{The $(k,0)$ Spectral Weight Analyses}
As argued above, very slow convergence is expected for 
the integrated $D=2$, $J=0+1$ $(k,0)$ spectral weight OPE series.
Such slow convergence is seen explicitly in the results reported in
Refs.~\cite{kmcwvus06,bck05}. The situation for $k=0$ is of particular
practical interest since the $s_0=m_\tau^2$, $(0,0)$ $us$
spectral integral is fixed by the total strangeness branching fraction.
Spectral integral errors can thus be reduced through improvements to
the various exclusive strange branching fractions. Such improvements are
much less experimentally challenging than would be an improved determination
of the full $us$ spectral distribution, which analyses employing other 
$s_0$ and/or other weights would require. The very small 
normalization of the $(0,0)$ $D=2$ OPE integral, which makes for a rather 
weak dependence of $V_{us}$ on $m_s$, is another favorable feature of this 
weight~\cite{pichetalvus}.
Unfortunately, these positive features must be weighed against
the very poor convergence of the integrated $D=2$ OPE series,
and the concomitant difficulty in obtaining a reliable
estimate of the truncation uncertainty. In the most recent version of this 
analysis (see the last of Refs.~\cite{pichetalvus}) $V_{us}$ is obtained
using external input for $m_s$, and $s_0=m_\tau^2$ only. The quoted 
{\it combined} OPE-induced uncertainty is $\pm 0.0011$.
As seen in Ref.~\cite{kmcwvus06}, however, the $(0,0)$ sum
rule displays rather poor $s_0$-stability. $V_{us}$, e.g., changes
by $0.0021$ even over the rather restricted range 
$m_\tau^2-0.4 \ {\rm GeV}^2<s_0<m_\tau^2$.
An alternate estimate of the $D=2$ truncation uncertainty, obtained
by taking {\it twice} the quadrature sum of the last term kept and the
correlator-minus-Adler-function difference yields a result,
$\pm 0.0022$, much more in keeping with the size of the observed $s_0$ 
instability. Unfortunately, since a reduction in the truncation
uncertainty does not appear likely, this more conservative
estimate puts a sub-$1\%$ $V_{us}$ determination out of reach
of the $(0,0)$ spectral weight analysis. Figure 1
shows the OPE and spectral integral differences, as a function of $s_0$, for
various input $m_s\equiv m_s(2\ {\rm GeV})$, with $V_{us}$ in each 
case obtained by fitting to the spectral integral difference at $s_0=m_\tau^2$.
The very different $s_0$-dependences of the two curves
is the source of the $s_0$-instability in the extracted $V_{us}$ values. 
The figure makes clear that the instability in $V_{us}$ found in
Ref.~\cite{kmcwvus06} is not specific to the $m_s$ employed as input
in that analysis. 

A final illustration of the problematic features of the $(k,0)$ 
spectral weight analyses is provided in Figure 4. Given any pair
of $(k,0)$ weights, $m_s$ and $V_{us}$ can be obtained by
a simultaneous fit using the two $s_0=m_\tau^2$ spectral integrals
as input. The figure shows the $1\sigma$ contours for a series of such fits.
It is evident that no good common fit region for
$m_s$ and $V_{us}$ exists, further strengthening
the conclusion that the OPE representations for the $(k,0)$ spectral
weights are not under sufficiently good control to allow
a reliable determination of $m_s$ and $V_{us}$.

\subsection{Non-Spectral Weight Analyses}
To reduce the problems associated with the slow convergence of the
$D=2$, $J=0+1$ OPE series, we shift to FESR's based on three
non-spectral weights, $w_{10}$, $\hat{w}_{10}$, and $w_{20}$, constructed
in Ref.~\cite{km00}. By design, these weights simultaneously (i) improve $D=2$
convergence; (ii) suppress spectral contributions above
$1\ {\rm GeV}^2$ (where $us$ errors are large); and (iii) control 
weight coefficients $c_m,\, m>2$ (which might otherwise enhance
$D>6$ OPE contributions). Ref.~\cite{km00} and Table I 
of Ref.~\cite{kmcwvus06} show explicitly the much improved 
$D=2$ convergence which results. Table II
of Ref.~\cite{kmcwvus06} (the left, ``ACO'', half corresponding to the 
$us$ data treatment used here) also shows the much improved stability of 
$V_{us}$ with respect to $s_0$ obtained for these weights, at least
for the PDG04 input $m_s(2\ {\rm GeV})=105\pm 25$
MeV employed there. Figures 2 and 3 show the $m_s$-dependent OPE vs.
spectral integral comparisons, analogous to those shown for the $(0,0)$ 
spectral weight in Figure 1, for $w_{20}$ and $w_{10}$, respectively.
The results for $\hat{w}_{10}$, which are similar, have been 
omitted for brevity, but may be found in Ref.~\cite{kmcwinprep}.
The existence of a window of $m_s$ values over which
improved $s_0$-stability for $V_{us}$ will be obtained for all three 
non-spectral weights is clear. Figure 5 shows the $1\sigma$ 
contours for the various pairwise fits and combined 3-fold fit involving
these weights. A good common fit region exists, in
sharp contrast to the situation for the $(k,0)$ spectral weights.
This strongly suggests that the lack of a good common fit region
for the $(k,0)$ spectral weights is a result of the poor
convergence behavior of the relevant integrated OPE series.

The results of Figure 5 correspond to $V_{us}=0.2202\pm 0.0046$ 
and $m_s\equiv m_s(2\ {\rm GeV})=89\pm 25$ MeV. A modest reduction of
the combined fit region can be achieved by adding a new non-spectral 
weight, $w_8(y)$, to the analysis~\cite{kmcwinprep}. An excellent common 
fit region remains. The central fit values and (somewhat) reduced 
errors are then~\cite{kmcwinprep}
\begin{equation}
V_{us}=0.2213\pm 0.0039,\ \ m_s=97\pm 19\ {\rm MeV}\ .
\label{results}\end{equation}
The result for $V_{us}$ is compatible, within errors, with both (i) the 
ICHEP06 $K_{\ell 3}$ review update~\cite{antonelli06}, 
$V_{us}=0.2232\pm 0.0006$ ($0.2249\pm 0.0019$ if the most
recent lattice input for $f_+(0)$~\cite{latticefplus06} 
is replaced by the old Leutwyler-Roos value~\cite{lr84}),
and (ii) the recent MILC update of the 
$\Gamma [K_{\mu 2}]/\Gamma [\pi_{\mu 2}]$ determination~\cite{milcvus},
$V_{us}=0.2223^{+0.0026}_{-0.0013}$. 
The result for $m_s$ is in excellent agreement with recent
strange scalar and PS sumrule, and $N_f=2+1$ lattice,
results.

To see how a reduction in the $us$ spectral errors might impact
the errors on $V_{us}$ and $m_s$, we consider a scenario in
which the $us$ spectral function central values remain unchanged 
but the errors (covariances) are reduced by a factor of $3$ ($9$).
The combined fit errors on $V_{us}$ and $m_s$ are reduced to
$\pm 0.0015$ and $\pm 13$ MeV, respectively. This, of course, provides
only a rough guide, since non-trivial
shifts in the central $us$ spectral function values are certainly to be
expected. Nonetheless, the exercise makes clear that very significant
reductions in the $V_{us}$ error should be anticipated once B factory
data becomes available. A more modest reduction is observed for the $m_s$
errors.
 


\begin{figure}
\unitlength1cm
\caption{$s_0$-stability plots for the $(0,0)$
spectral weight} 
\begin{minipage}[t]{10.0cm}
\begin{picture}(9.9,13.5)
\epsfig{figure=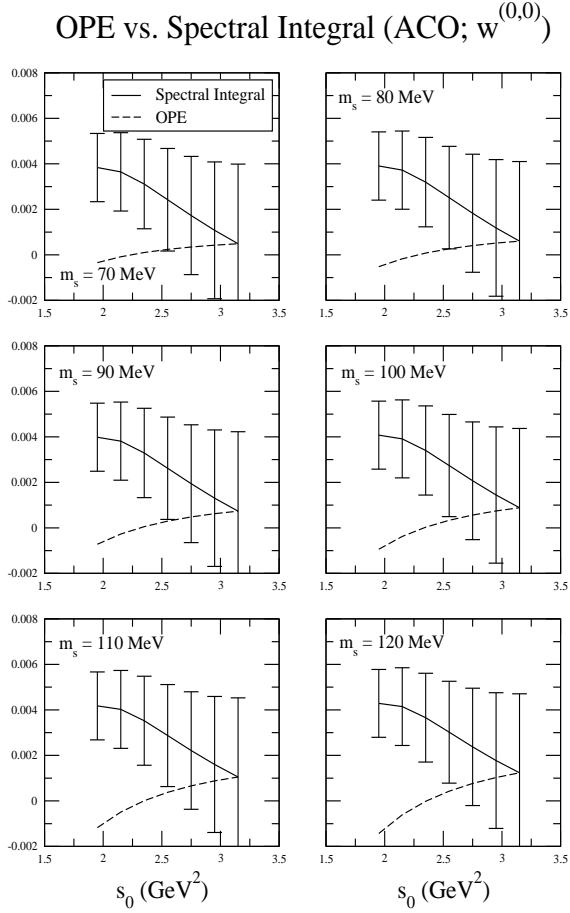,height=13.4cm,width=9.8cm}
\end{picture}\qquad\quad
\end{minipage}
\label{w00msrange}
\end{figure}

\begin{figure}
\unitlength1cm
\caption{$s_0$-stability plots for $w_{20}$}
\begin{minipage}[t]{10cm}
\begin{picture}(9.9,13.5)
\epsfig{figure=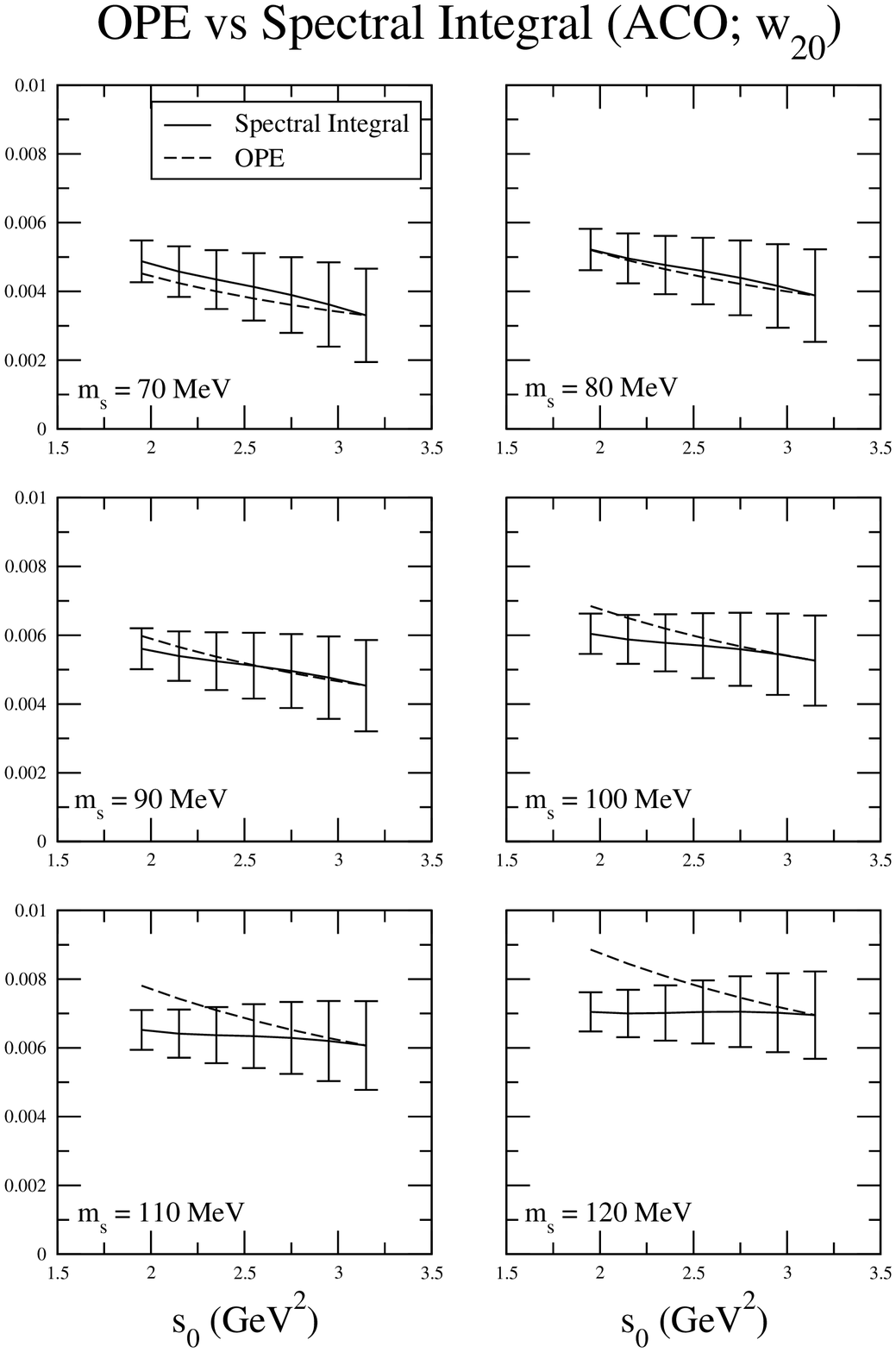,height=13.4cm,width=9.8cm}
\end{picture}\qquad\quad
\end{minipage}
\label{w20msrange}\end{figure}


\begin{figure}
\unitlength1cm
\caption{$s_0$-stability plots for $w_{10}$}
\begin{minipage}[t]{10cm}
\begin{picture}(9.9,13.5)
\epsfig{figure=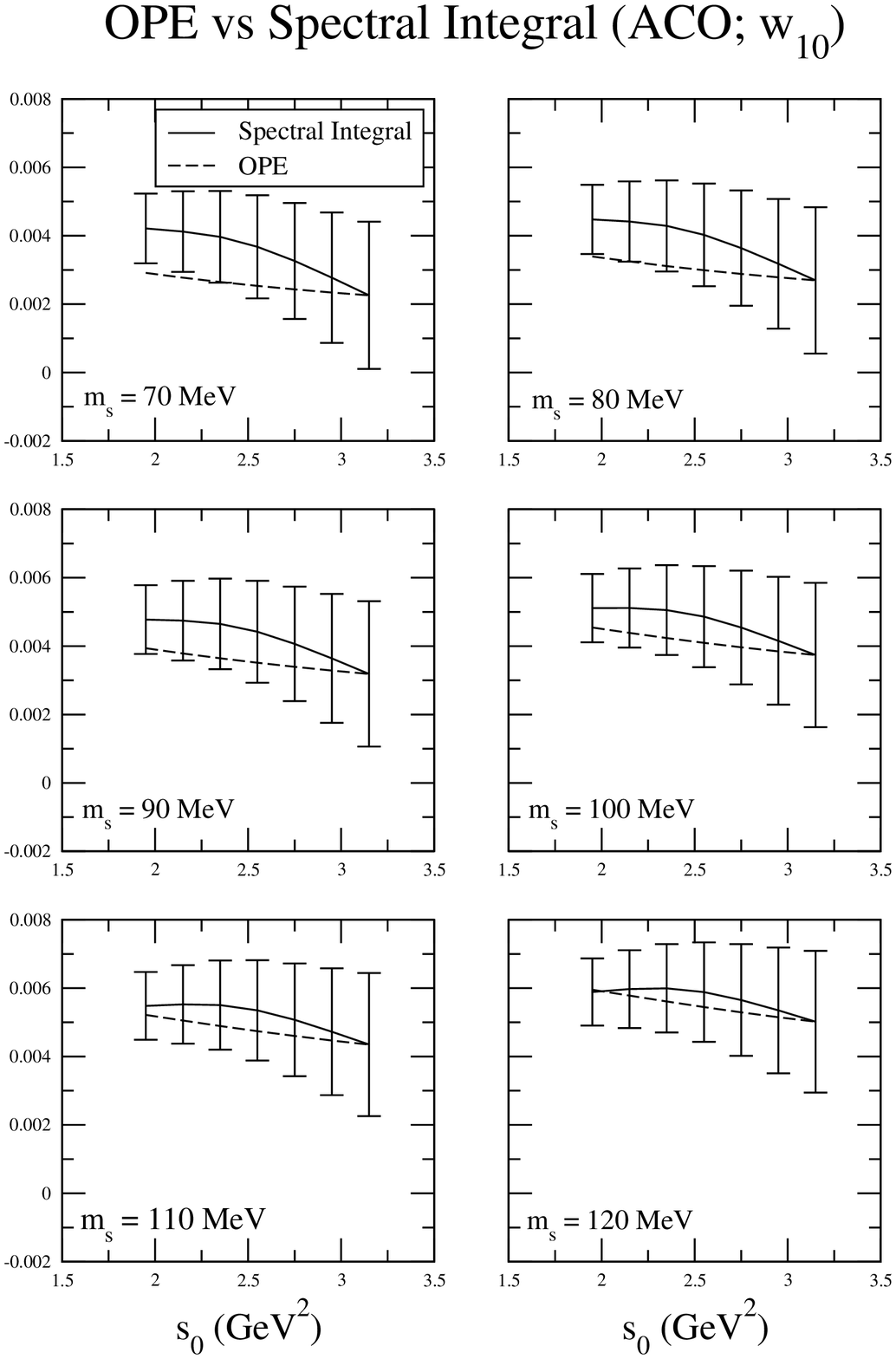,height=13.4cm,width=9.8cm}
\end{picture}\qquad\quad
\end{minipage}
\label{w10msrange}\end{figure}


\begin{figure}
\unitlength1cm
\caption{$(k,0)$ spectral weight joint fit contours}
\rotatebox{270}{\mbox{
\begin{minipage}[t]{8cm}
\begin{picture}(7.9,7.9)
\epsfig{figure=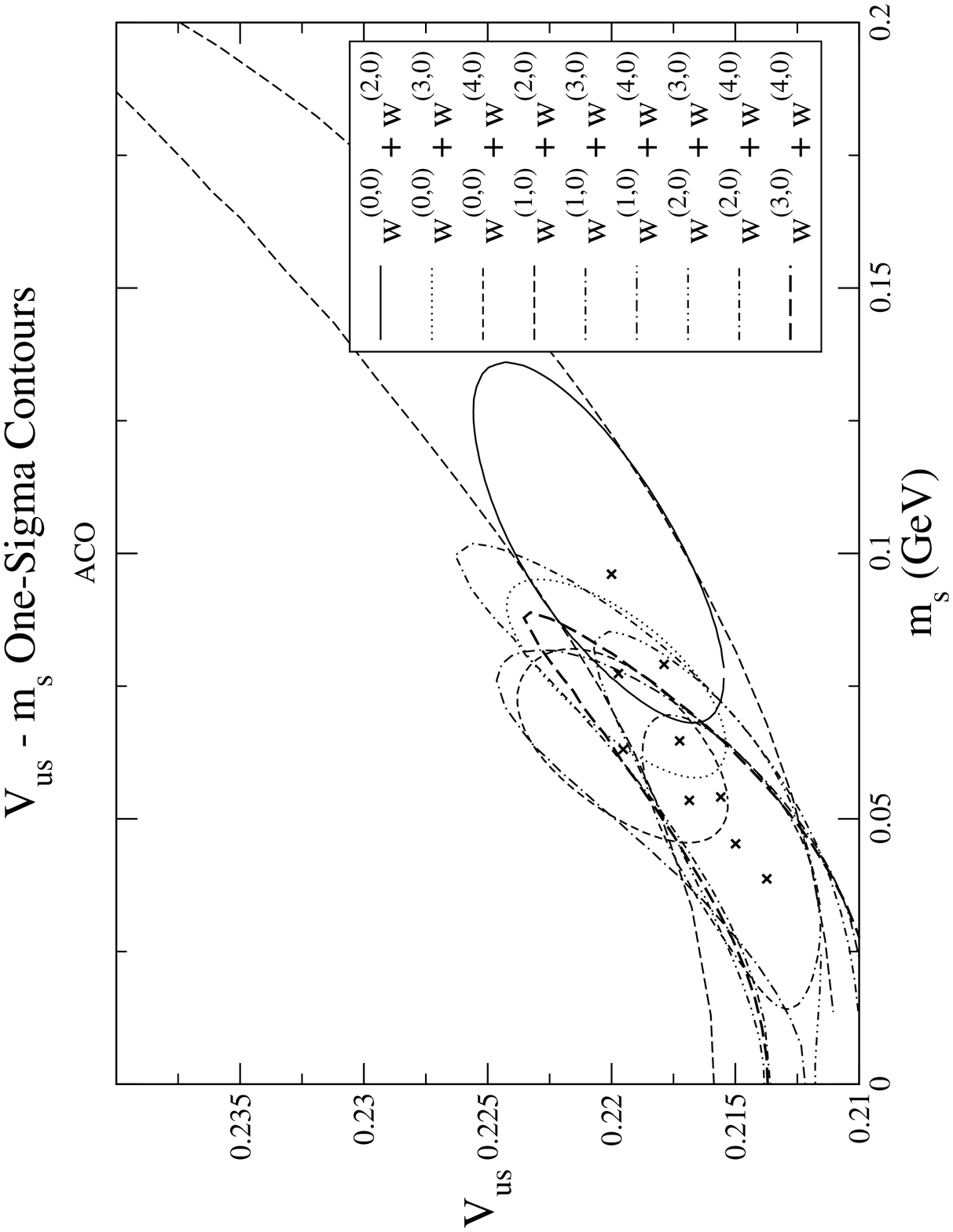,height=7.8cm,width=7.8cm}
\end{picture}
\end{minipage}}}
\label{wspecwtcontours}\end{figure}

\begin{figure}
\unitlength1cm
\caption{Non-spectral weight joint fit contours}
\rotatebox{270}{\mbox{
\begin{minipage}[t]{8cm}
\begin{picture}(7.9,7.9)
\epsfig{figure=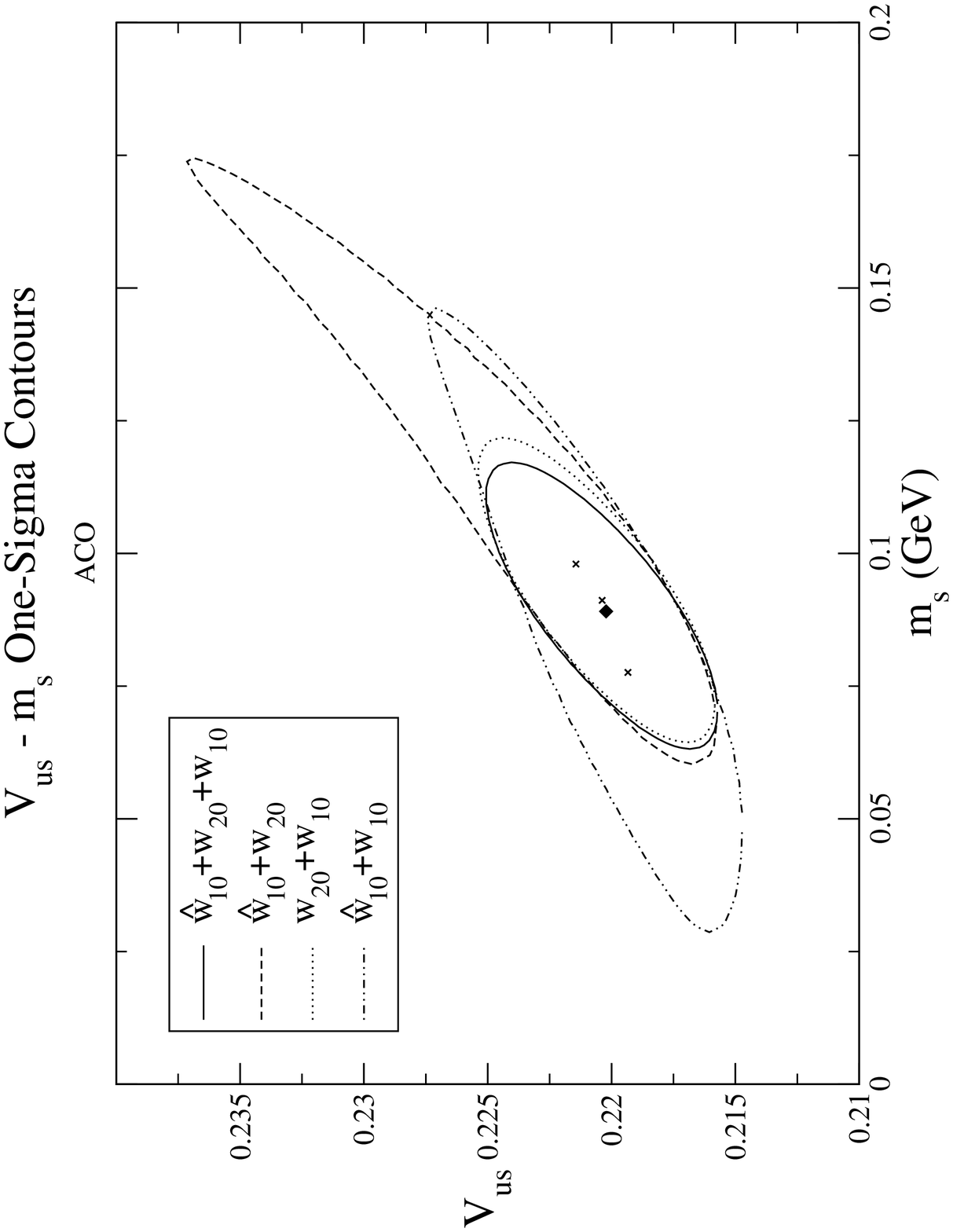,height=7.8cm,width=7.8cm}
\end{picture}
\end{minipage}}}
\label{nonspecwtcontours}\end{figure}




\begin{thebibliography}{99}

\bibitem{bnpetc}E. Braaten, S. Narison and A. Pich, Nucl. Phys. {\bf B373}
(1992) 581; see also A. Pich, Nucl. Phys. Proc. Suppl. {\bf 39BC} (1995) 326
for a review and list of earlier references.
\bibitem{taumsrefs}See, e.g., K.G. Chetyrkin and A. Kwiatkowski,  Z. Phys. 
{\bf C59} (1993) 525 and hep-ph/9805232; K. Maltman, Phys. Rev. D58 (1998) 
093015; 
K.G. Chetyrkin, J.H. Kuhn and A.A. Pivovarov, Nucl. Phys. {\bf B533} (1998) 
473; A. Pich and J. Prades, JHEP {\bf 06} (1998) 013, Nucl. Phys. Proc. 
Suppl. 74 (1999) 309 and JHEP 9910 (1999) 004; J.G. K\"orner, F. Krajewski 
and A.A. Pivovarov, Eur. Phys. J {\bf C20} (2001) 259 and {\it ibid.}
{\bf C14} (2000) 123.
\bibitem{km00}J. Kambor and K. Maltman, Phys. Rev. {\bf D62} (2000) 093023. 
\bibitem{longposviol}K. Maltman and J. Kambor, Phys. Rev. {\bf D64} (2001)
093014. 
\bibitem{pichetalvus}E. Gamiz {\it et al.}, JHEP {\bf 0301} (2003) 060;
Phys. Rev. Lett. {\bf 94} (2005) 011803; hep-ph/0610246.
\bibitem{kmtau0204}K. Maltman, Nucl. Phys. Proc. Suppl. {\bf 123} (2003) 123;
{\it ibid.} {\bf 144} (2005) 65.

\bibitem{kmcwvus06}K. Maltman and C.E. Wolfe, Phys. Lett. {\bf B639}
(2006) 283.

\bibitem{cleoud}T. Coan {\it et al.} (The CLEO Collaboration),
Phys.Lett. {\bf B356}, 580 (1995); S. Anderson {\it et al.} 
(The CLEO Collaboration), Phys. Rev. {\bf D61} (2000) 112002.
\bibitem{alephud}R. Barate {\it et al.} (The ALEPH Collaboration), 
Z. Phys. {\bf C76} (1997) 15; Eur. Phys. J. {\bf C4} (1998) 409.
\bibitem{alephfinalspec}S. Schael {\it et al.} (The ALEPH Collaboration),
Phys. Rep. {\bf 421} (2005) 121.
\bibitem{opalud}K. Ackerstaff {\it et al.} (The OPAL Collaboration),
Eur. Phys. J. {\bf C7}, 571 (1999).
\bibitem{alephus99}R. Barate {\it et al.} (The ALEPH Collaboration),
Eur. Phys. J. {\bf C11}, 599 (1999).
\bibitem{cleous0305}R.A. Briere {\it et al.} (The CLEO Collaboration),
Phys. Rev. Lett. {\bf 90}, 181802 (2003);
K. Arms {\it et al.} (The CLEO Collaboration), Phys.
Rev. Lett. {\bf 94}, 241802 (2005). 
\bibitem{opalus04}G. Abbiendi {\it et al.} (The OPAL Collaboration),
Eur. Phys. J. {\bf C35}, 437 (2004).
\bibitem{bck05}P.A. Baikov, K.G. Chetyrkin and J.H. Kuhn,
Phys. Rev. Lett. {\bf 95}, 012003 (2005).
\bibitem{longconv}K. Maltman, Phys. Rev. D58 (1998) 093015; 
K.G. Chetyrkin and A. Kwiatkowski,  Z. Phys. {\bf C59} (1993) 525 and 
hep-ph/9805232; A. Pich and J. Prades, JHEP {\bf 9806} (1998) 013.

\bibitem{longposviolfootnote}The results for $m_s$ from earlier inclusive 
$(k,0)$ spectral weight analyses, in which the truncated longitudinal OPE 
representation was employed, display a very strong unphysical $k$-dependence,
a large part of which turns out to be a consequence of the implicit
violation of spectral positivity in those analyses~\cite{longposviol}. 
\bibitem{mkps}K. Maltman and J. Kambor, Phys. Rev. {\bf D65}, 074013
(2002); Phys. Lett. {\bf B517} (2001) 332.
\bibitem{jopss}M. Jamin, J.A. Oller and A. Pich, Nucl. Phys. {\bf B587}
(2000) 331; {\it ibid.} {\bf B622} (2002) 279; hep-ph/0605095.
\bibitem{msssnew}M. Jamin, J.A. Oller and A. Pich, Phys. Rev. 
{\bf D74} (2006) 074009.
\bibitem{sumrulems}P.A. Baikov, K.G. Chetyrkin and J.H. Kuhn, Phys. Rev. Lett.
{\bf 96} (2006) 012003; K.G. Chetyrkin and A. Khodjamirian, hep-ph/0512295; 
M. Jamin, J.A. Oller and A. Pich, Eur. Phys. J. {\bf C24} (2002) 237.
\bibitem{mslattice}Q. Mason, {\it et al.} (The HPQCD Collaboration),
Phys. Rev. {\bf D74} (2006) 114501; C. Bernard {\it et al.} (The MILC
Collaboration), hep-lat/0609053; T. Ishikawa {\it et al.} (The CP-PACS
and JLQCD Collaborations), hep-lat/0610050. 
\bibitem{bckzin04}P.A. Baikov, K.G. Chetyrkin and J.H. Kuhn, Nucl.
Phys. B Proc. Suppl. {\bf 135} (2004) 243.
\bibitem{pdg06}W.-M. Yao {\it et al.} (The Particle Data Group),
J. Phys. G {\bf 33} (2006) 1.
\bibitem{cks97}K.G. Chetyrkin, B.A. Kniehl and M. Steinhauser, Phys.
Rev. Lett. {\bf 79} (1997) 2184.
\bibitem{betagamma4}T. van Ritbergen, J.A.M. Vermaseren and S.A. Larin,
Phys. Lett. {\bf B400} (1997) 397 and {\it ibid.} {\bf B405} (1997) 327;
K.G. Chetyrkin, Phys. Lett. {\bf B404} (1997) 161;
M. Czakon, Nucl. Phys. {\bf B710} (2005) 485.

\bibitem{footnote1}Readers familiar with the smallness of
the $O(\bar{a}^2)$ term of the integrated contour-improved $(0,0)$
spectral weight series, and the similar smallness of the
$O(\bar{a}^3)$ (respectively $O(\bar{a}^4)$) term of the
corresponding $(1,0)$ (respectively $(2,0)$) series, might
be puzzled by this claim. The smallness of these terms,
however, results from close cancellations among contributions
from different parts of the integration contour. Such cancellations
are ``accidental'', in the sense that they occur only for 
one particular ($k$-dependent) power of $\alpha_s$, and do
not persist to higher orders. The growth with
$k$ of the power for which this closest cancellation occurs
is a result of the growth of the concentration of the 
support for the $(k,0)$ weight in the spacelike region.
A misleading impression of the convergence of the truncated series
can result if the truncation order happens to coincide
with the order of closest cancellation. The danger
posed by such a situation is one of the important arguments in
favor of performing the $s_0$-stability tests advocated in the text.
\bibitem{kmfesr}K. Maltman, Phys. Lett. {\bf B440} (1998) 367
Nucl. Phys. Proc. Suppl. {\bf 123} (2003) 123.
\bibitem{cdgmudvma}V. Cirigliano, J.F. Donoghue, E. Golowich and 
K. Maltman, Phys. Lett. {\bf B555} (2003) 71; 
V. Cirigliano, E. Golowich and K. Maltman, Phys. Rev.
{\bf D68} (2003) 054013.
\bibitem{kmcwinprep}K. Maltman and C.E. Wolfe, ``Joint Extraction
of $m_s$ and $V_{us}$ from Hadronic $\tau$ Decay Data'', 
in preparation
\bibitem{ciptbasic}A.A. Pivovarov, Sov. J. Nucl. Phys. {\bf 54} (1991) 676,
Z. Phys. {\bf C53} (1992) 461; F. Le Diberder and A. Pich,
Phys. Lett. {\bf B286} (1992) 147, {\it ibid.} {\bf B289} (1992) 165.
\bibitem{chen01}S. Chen {\it et al.}, Eur. Phys. J. {\bf C22} (2001) 31.
\bibitem{antonelli06}M. Antonelli, hep-ex/0610070
\bibitem{latticefplus06}D. J. Antonio {\it et al.}, hep-lat/0610080.
\bibitem{lr84}H. Leutwyler and M. Roos, Z. Phys. {\bf C25} (1984) 91.
\bibitem{milcvus}C. Bernard {\it et al.} (The MILC Collaboration), 
hep-lat/0609053. 



\end{thebibliography}
\end{document}